2016 IUTAM Symposium on Nanoscale Physical Mechanics

# Graphene and Carbon Nanotube Hybrid Structure: A Review

## Kang Xia, Haifei Zhan, and Yuantong Gu*

*School of Chemistry, Physics and Mechanical Engineering, Queensland University of Technology (QUT), Brisbane QLD 4001, Australia*

**Abstract**

Graphene has been reported with record-breaking properties which have opened up huge potential applications. Considerable amount of researches have been devoted to manipulate or modify the properties of graphene to target a more smart nanoscale device. Graphene and carbon nanotube hybrid structure (GNHS) is one of the promising graphene derivatives. The synthesis process and the mechanical properties are essential for the GNHS based devices. Therefore, this review will summarize the recent progress of the highly ordered GNHS synthesis/assembly, and discuss the mechanical properties/behaviors of GNHS under various conditions as obtained from molecular dynamics simulations.
© 2017 The Authors. Published by Elsevier B.V.
Peer-review under responsibility of organizing committee of Institute of the 2016 IUTAM Symposium on Nanoscale Physical Mechanics.

## 1. Introduction

The excellent mechanical, physical, and chemical properties of low-dimensional carbon materials (e.g., graphene, carbon nanotube - CNT) have enabled them as promising building blocks for three-dimensional nanoarchitectures. In this regard, extensive interests have been attracted to synthesize graphene and carbon nanotube hybrid structures[1]. Earlier works suggested that the CNT-graphene structure is a promising solution for quick energy dissipation, which could enhance interface thermal conductivity of the 'building blocks' for future nanoscale mechanical and electrical devices[2]. In the electricity perspective, Frederico et al.[3] and Yu et al.[4] found the 'CNT pillared-graphene' system would extend the excellent electrical conductivity of graphene and nanotube to three dimensions. According to Fan et al.[5], the double layer configuration of CNT and graphene hybrid is supposed to endow the structure with better electrochemical performance, which indicates that the hybrid structure is suitable for electrode as used supercapacitors. Further, CNT-graphene hybrid structure is also found to have huge potential in hydrogen storage (hydrogen storage capacity can be as high as 41 gL[-1]), due to its extremely large surface area[6]. This review aims to briefly summarize the latest progress in ordered GNHS synthesis, and also the understanding of the mechanical behaviors of GNHSs under various loading conditions, which will shed lights on their design and engineering implementations.

* Corresponding author. Tel.: +61731381009
*E-mail address:* yuantong.gu@qut.edu.au

## 2. Synthesis and assemblies of GNHS

Early researches on GNHS family compounds are focused on the assembly of GNHS with ordered structure for specific application. Generally, the synthesis/fabrication approaches of GNHSs can be classified into four different categories, including solution processing/casting[7-12], layer-by-layer deposition[13-15], vacuum filtration[16,17] and chemical vapour deposition (CVD)[5,18-31]. Among all these methods, CVD approach is shown to build hierarchical nanostructures with reasonable structure stability and mechanical strength[18,32].

In 2012, Yu et al disclosed an approach to synthesise seamless, covalently bonded three-dimensional graphene and carbon nanotube hybrid material with the aid of a floating buffer layer in chemical vapour deposition environment[4], as illustrated in Fig. 1. In detail, graphene is synthesised on the copper foil first, then iron catalyst and alumina buffer layer are deposited on top of graphene respectively using electron evaporation. In the growth stage the buffer is lifted up and CNT carpet grow directly out of the graphene. The density of CNTs forest can be well controlled by the thickness of the iron catalyst layer and the growth rate can be as high as 120 μm in 10 mins. The STEM reveal that the graphene and CNT junction region are well covalent bonded and are suggested to be utilized in the high performance supercapacitor and energy storages.

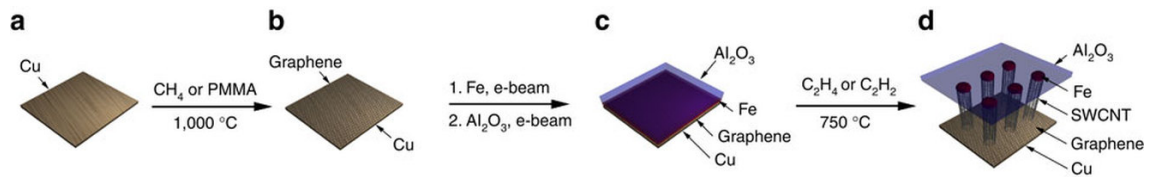

Fig. 1. Scheme for the synthesis of CNT carpets directly from graphene[4]. (a) Copper foil substrate. (b) Graphene is formed on the copper foil by CVD or solid carbon-source growth. (c) Iron and alumina are deposited on the graphene-covered copper foil by using e-beam evaporation. (d) A CNT carpet is directly grown from graphene surface. The iron catalyst and alumina protective layer are lifted up by the CNT carpet as it grows. Reprinted by permission from Macmillan Publisher Ltd: Nature Communications, copyright 2016.

To facilitate the application of GNHS in the field of energy storage, Cheng et al.[18] developed a two-step CVD growth method (see Fig. 2). First of all, FeMo/vermiculite composed of exfoliated vermiculite (EV) embedded with FeMo nanoparticles in the interlayer space serves as the bifunctional catalyst to help the synthesis of CNT/graphene in a low temperature CVD environment of 650 °C. Subsequently, a higher temperature of 950 °C CVD was conducted for uniform graphene layer deposition. To obtain the high carbon purity GNHS, facile acid treatments were applied to remove the FeMo/vermiculite catalysts. The obtained GNHS demonstrates outstanding compression and recovery performance, which shows the potential in the energy storage industry.

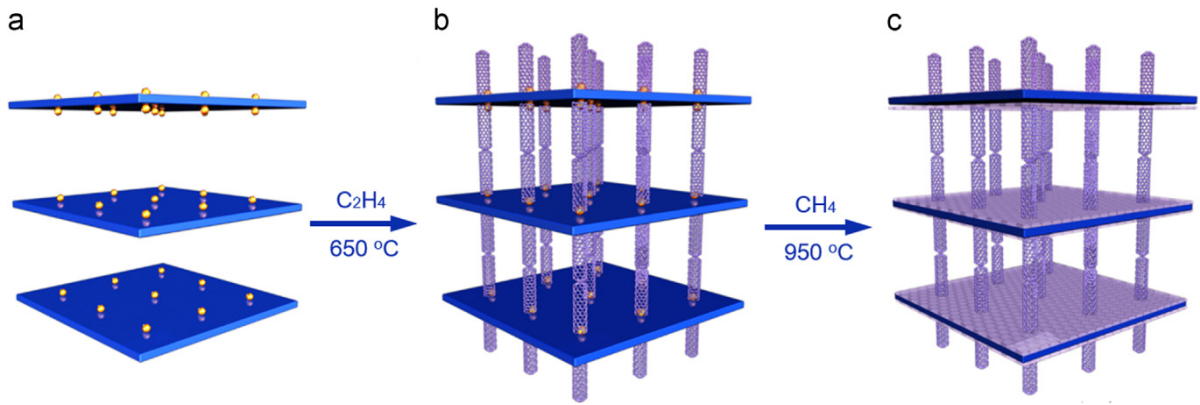

Fig. 2. Schematic illustration for the synthesis of aligned GNHS on bifunctional catalysts. (a) Metal NPs nucleated on the surface of EV flakes; (b) aligned CNT grown on EV flakes by a low temperature CVD; (c) graphene sheets deposited on EV flake by high temperature CVD for aligned GNHS formation. Reprinted with permission from reference[18]. Copyright 2016 Elsevier.

## 3. Mechanical behaviors of GNHS

A comprehensive understanding of the mechanical properties of GNHS is a crucial prerequisite for its engineering applications, which relies heavily on the experimental tests and numerical simulations. In 2014, Cheng et al.[18] performed compression tests on a freestanding block of aligned GNHS with the size of 4 mm and a CNT length of ~ 10 μm was repeatedly tested at strain range from 10 to 95%. It is found that the roots of aligned which anchored on the graphene sheets provides excellent loading transfer between graphene and CNTs at high stress loading and the compressive modulus can reach as high as 2.3 GPa before the collapse of the structure. The GNHS shows excellent structure stability and energy density of 237.1 kJ/kg at 95% strain and exhibits superiority over other conventional material like steel (as compared in Fig. 3).

In terms of numerical simulations, several works have been reported on GNHS, focusing on different length scales. Based on structural molecular mechanics approach, Sangwook et al.[33] conducted a finite element analysis to predict the effective mechanical stiffness properties of GNHSs. This parametric study suggests that the pillar length and inter-pillar distance significantly influence Young's modulus. That is, GNHSs with shorter CNTs yield higher planar Young's and shear moduli, while those with smaller inter pillar distance yield higher through-thickness moduli. Negative in-plane Poisson's ratios are observed for all sets of GNHSs and are believed to be associated with curvature at the junction. Xu et al.[34] performed a thorough investigation on GNHSs using MD approach, Young's moduli obtained from a tensile loading test along the armchair direction of the graphene is found to be around 0.9 TPa, and the corresponding tensile strength and strain are ~ 65-72 GPa and 0.09, respectively. These properties are significantly smaller compared with that of the pristine graphene. Similar trends are also observed in our recent works[35,36]. Generally, the GNHS exhibits a brittle tensile behaviour and the stretched carbon bonds are observed around the connecting regions between graphene and CNT during the elastic deformation period. Affluent formations of monoatomic chains and rings are found at the fracture vicinity (Fig. 4).

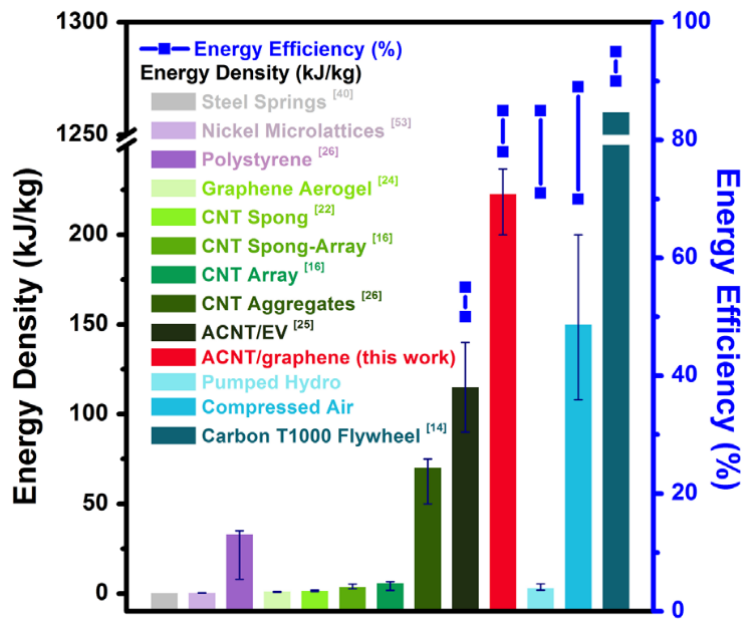

Fig. 3. Energy density of various mechanical energy storage materials. Reprinted with permission from reference[18]. Copyright 2016 Elsevier.

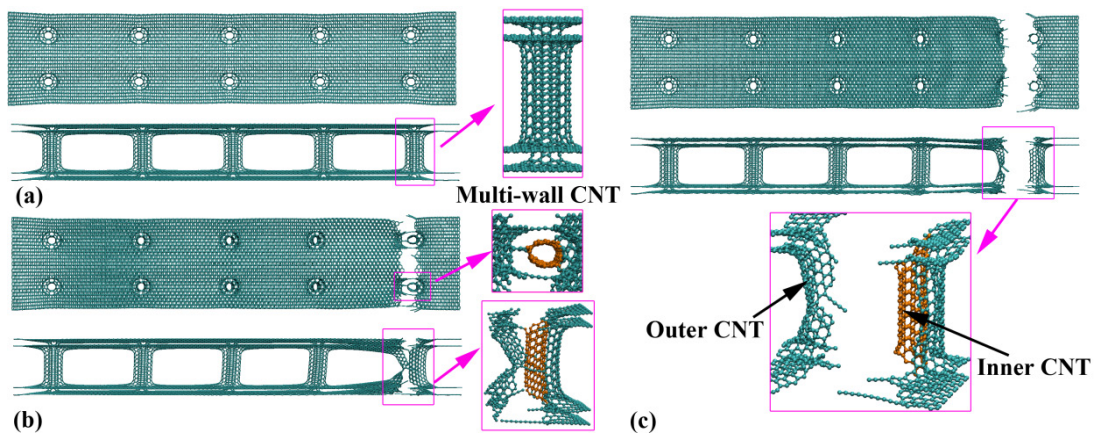

Fig. 4. GNHS with multi-walled CNTs at different strain of: (a) 9.4%, inset shows a multi-walled CNT-graphene junction; (b) 10.1%, inset shows the deformation around the multi-walled CNT-graphene junctions; (c) 10.15%, inset shows the fracture of the outer CNTs. Reprinted with permission from reference[35].

To facilitate different application purposes, physical or chemical modification is frequently applied. Doping is one of the effective approaches that allows for the intrinsic modification to the electrical and chemical properties of nanomaterials, such as graphene[37]. In this regard, our recent work has assessed how the dopants would affect the mechanical properties of GNHS[38] through large-scale molecular dynamics simulations. It is found that with the presence of dopants, the hybrid structures usually exhibit lower yield strength, Young's modulus, and earlier yielding compared with that of a pristine hybrid structure. Interestingly, increase the dopant density (below 2.5%), significant degradation in Young's modulus or yield strength is not observed (Fig. 5).

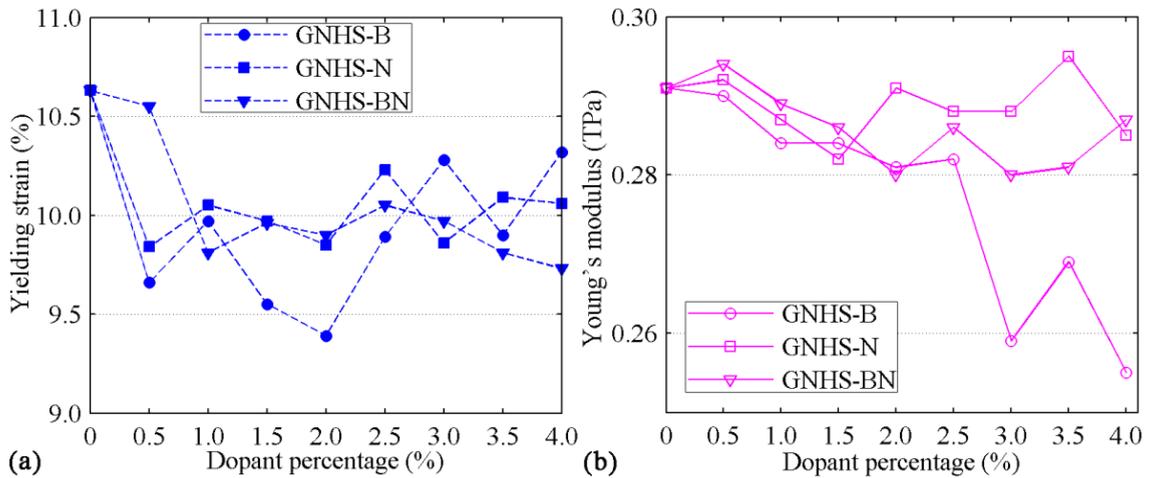

Fig. 5. Yielding strain and Young's modulus as a function of the density of N-, B-, and NB-dopant. Reprinted with permission from reference[38]. Copyright 2016 Beilstein Journal of Nanotechnology.

Similar as obtained from the pristine GNHS, the failure of the doped GNHS is found to initiate at the region where the nanotube and graphene sheets are connected. After failure, monatomic chains are normally observed around the failure region. Dangling graphene layers without the separation of a residual CNT wall are found to adhere to each other after failure with a distance of about 3.4 Å. Such deformation mechanisms are uniformly observed from the boron, nitrogen, or both-doped GNHS.

In order to fully understand the mechanical properties and deformation pattern of GNHS under more complex loading condition, MD based nanoindentation test is carried out by Wang et al[39]. For the samples studied, a strong correlation is found between CNT dimeter and the mechanical strength. For instance, vary the dimeter of CNT, a maximum increase in hardness of approximately 40% is observed, also, the reduced modulus increases with increasing CNT diameter during the indentation process. In addition, the maximum load which observed before the collapse of the GHNS decreases with temperature, when the temperature is 700 K the GNHS bears the smallest load which is ~3.0 nN (Fig. 7), this observation probably originates from the increased atomic mobility at higher temperatures. The loading and unloading does not match perfectly with each other which indicates viscoelastic characteristic.

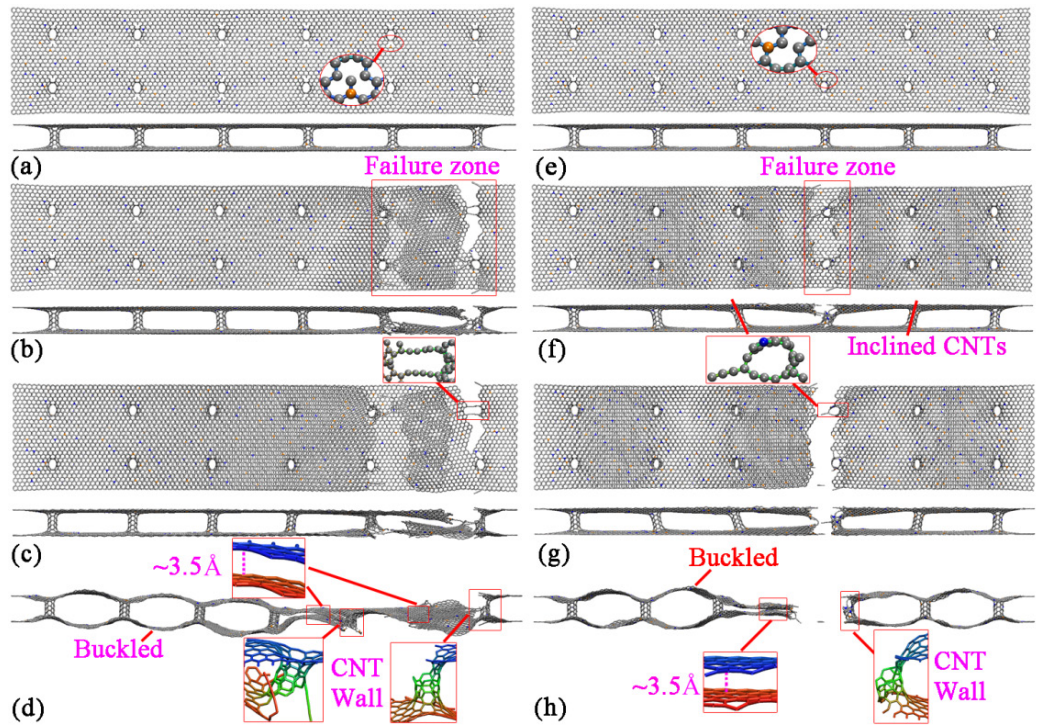

Fig. 6. Atomic configurations of GNHS-0.75%B0.75%N at different strain of: (a) 0.097; (b) 0.101 (c) @ 0.102; (d) 0.115. Atomic configurations of GNHS-1.5%B1.5%N at the strain of: (e) 0.097; (f) 0.101; (g) 0.102; (h) 0.103. Reprinted with permission from reference[38]. Copyright 2016 Beilstein Journal of Nanotechnology.

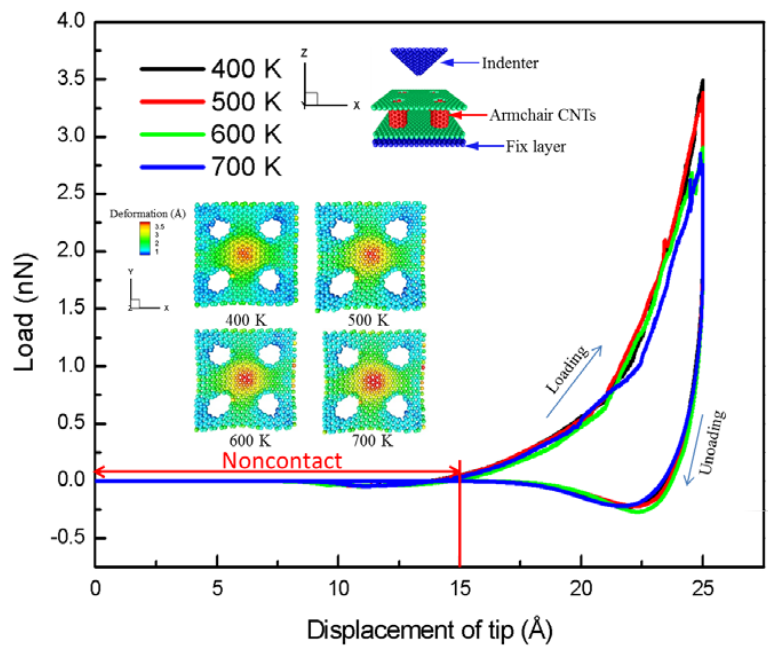

Fig. 7. Variation of load force of zigzag 3D-pillared graphene with tip displacement at 400, 500, 600, and 700K (in dentation depth was set at 1.0 nm). Reprinted with permission from reference[39]. Copyright 2016 Cambridge University Press and Copyright Clearance Center.

## 4. Summary

Graphene and carbon nanotube hybrid structure is one of the promising graphene derivatives, which have great potential applications in areas such as energy storage, supercapacitor and building blocks composite. Experimental works have shown that nicely aligned hybrid structures between graphene and carbon nanotube are achievable. Their highly tailorable structures endow them with highly tunable mechanical properties.

It is worth mentioning that other properties of the graphene and CNT hybrid structures are also of great interest, such as chemical properties and thermal transport properties. In this regard, Varshney et al.[2] find that the thermal transport is governed by the minimum interpillar distance and the CNT length. Additionally, plenty of works have also been reported on other carbon nanoarchitectures, such as CNT based hexagonal structure[40] or super-tube structure[41]. The large variety of hybrid structures as constructed from low dimensional carbon materials (graphene/CNT), signifying their promising and wide applications in engineering sectors.

## Acknowledgement

Support from the ARC Discovery Project (DP150100828) is gratefully acknowledged.